\documentclass[twocolumn,amsmath,amssymb,prd,nofootinbib]{revtex4}

\def\al{\alpha}
\def\be{\beta}
\def\ga{\gamma}
\def\de{\delta}

\def\et{\eta}

\def\ka{\kappa}
\def\la{\lambda}

\def\rh{\rho}

\def\si{\sigma}

\def\ta{\tau}

\def\ph{\phi}

\def\ch{\chi}

\def\om{\omega}

\def\Om{\Omega}

\def\cG{{\cal G}}

\def\cL{{\cal L}}

\def\fr#1#2{{{#1} \over {#2}}}

\def\frac#1#2{{\textstyle{{#1}\over {#2}}}}

\def\prt{\partial}

\def\etal{{\it et al.}}
\def\pt#1{\phantom{#1}}

\newcommand{\beq}{\begin{equation}}
\newcommand{\eeq}{\end{equation}}
\newcommand{\bea}{\begin{eqnarray}}
\newcommand{\eea}{\end{eqnarray}}
\newcommand{\bit}{\begin{itemize}}
\newcommand{\eit}{\end{itemize}}
\newcommand{\rf}[1]{(\ref{#1})}
\newcommand{\nn}{\nonumber}

\def\mn{{\mu\nu}}
\def\sb{\overline{s}}
\def\tb{\overline{t}}
\def\cb{\overline{c}}
\def\ab{\overline{a}}

\def\K5{{K_{jklm}}}
\def\hr{{\hat R}}
\def\hb{{\hat b}}
\def\pb{\overline {p}}

\begin{document}

\title{Velocity-dependent inverse cubic force and solar system gravity tests}

\author{Quentin G.\ Bailey}
\affiliation{Physics Department, Embry-Riddle Aeronautical University,
Prescott, AZ 86301, U.S.A.}

\author{Daniel Havert}
\affiliation{Physics Department, Indiana University,
Bloomington, Indiana 47405, U.S.A.}

\date{Sept 21, 2017}

\begin{abstract}

Higher mass dimension terms in an effective field theory framework for tests of spacetime symmetries are studied.
Using a post-Newtonian expansion method, 
we derive the spacetime metric and the equations of motion for a binary system.
This reveals an effective inverse cubic force correction to post-Newtonian General Relativity 
that depends on the velocity of the bodies in the system.
The results are studied in the context of laboratory and space-based tests including the effects on solar-system ephemeris, 
laser ranging observations, and gravimeter tests.
This work reveals the coefficient combinations for mass dimension $5$ operators controlling CPT violation for gravity that
can be measured using analysis from these tests.
Other tests including light propagation can be used to probe these coefficients.
Sensitivity estimates are provided and the results are contrasted 
with the minimal mass dimension $4$ terms in the gravity sector.

\end{abstract}

\maketitle

\section{Introduction}
\label{intro}

So far, 
General Relativity (GR) satisfies all experimental and observational tests.
Nonetheless there remains widespread interest in continuing to test foundational principles of GR, 
like the Einstein Equivalence Principle, 
which includes local Lorentz symmetry.
Furthermore, 
it is interesting to consider the role in GR of other fundamental spacetime symmetries 
that play a vital role in particle physics, 
such as the combined Charge, Parity, 
and Time-reversal symmetry, 
or CPT symmetry.

To test CPT and Lorentz symmetry requires a consistent test framework
that allows for violations of this symmetry in a generic way.
Such a framework is provided by the Standard-Model Extension (SME) \cite{ck9798, k04}.
The underlying hypothesis is that spacetime-symmetry breaking is generically described by the presence
of background tensor fields called the coefficients for Lorentz violation that couple to known fields.
This framework is constructed with effective field theory, 
and its action includes GR and the Standard Model of particle physics
plus a series of terms describing generic spacetime symmetry breaking for existing fields.
The symmetry-breaking terms are constructed out of all the possible scalars formed from the background
fields coupled to operators involving known fields describing matter or gravity.
The SME framework has been widely used for analyzing experimental and observational searches
for Lorentz and CPT violation \cite{tables}.

For local fields in flat spacetime, 
a breaking of CPT symmetry is necessarily accompanied by a breaking of Lorentz symmetry, 
as established in the anti-CPT theorem \cite{owg02}.
Issues involving discrete symmetries in GR have already been investigated in several works \cite{parity}.
CPT violation in gravity, 
stemming from the general construction of the SME, 
has been investigated in the context of gravitational waves but not yet in other tests \cite{km16}.
Note that the role of CPT symmetry in curved spacetime is not settled, 
and we focus in this work primarily on its role in linearized gravity 
on a flat background.
It is known that CPT-breaking effects cancel from the Newtonian gravitational potential 
and therefore do not appear dominantly in short-range gravity tests \cite{bkx15, km17}.
We show in this work that when considering the next order in powers of $v/c$ in a post-Newtonian expansion 
of the spacetime metric around a flat background, 
terms proportional to the fluid velocity arise and lead to subtle but potentially measurable effects 
for CPT violation.

A great deal of work exists in the literature on the broad topic of possible spacetime-symmetry violation in nature
and we do not summarize it here.  
The reader is referred to the many review articles that exist on the topic, 
in particular more recent ones in Refs.\ \cite{reviews}.
Furthermore, 
the action-based approach of the SME presented here is complementary to and overlaps with other approaches
such as metric-based test frameworks \cite{tegp} and specific models \cite{yunes1416}.

This paper is organized as follows.
In Sec.\ \ref{theory} of this paper, 
we discuss the Lagrange densities for the gravitational sector of the SME, 
and derive the field equations.
The post-Newtonian metric is derived and 
discussed in Sec.\ \ref{pnexpansion}, 
along with the associated two-body equations of motion.
The remainder of the paper is devoted to calculating observables in specific tests in Sec.\ \ref{tests}, 
with subsections on secular orbit changes, 
laser ranging tests, 
light propagation effects, 
and Earth laboratory tests.
Finally we summarize the results in Sec.\ \ref{discussion}, 
including a table of estimated sensitivities.
Throughout the paper we use natural units where $c=\hbar=1$ 
and we adopt where possible the conventions of previous works \cite{k04,bk06}.
We will also make use of abbreviations for multiple partial derivatives so that,  
$\prt_{jkl...}=\prt_j \prt_k \prt_l...$ and we will use parenthesis (brackets) 
for symmetrization (antisymmetrization) of indices with a factor of one half.

\section{Theory}
\label{theory}

The general setting for the SME treatment of spacetime-symmetry breaking is Riemann-Cartan spacetime, 
which includes torsion couplings.
One begins with a coordinate invariant set of scalars in the lagrange density added to General Relativity and the matter sector of the Standard Model.  
These extra terms break the spacetime symmetries of local Lorentz symmetry, 
diffeomorphism symmetry, 
and can also break CPT symmetry \cite{k04}.
The origin of the symmetry-breaking terms can be explicit or through a dynamical mechanism
such as spontaneous spacetime-symmetry breaking which may occur in an underlying theory.
While general spacetime settings with torsion and even nonmetricity have recently been studied \cite{tnm}, 
in this work we shall focus on the Riemann spacetime limit with vanishing torsion.

At present there are two approaches to the gravitational sector of the SME.
The first is a general coordinate invariant version where the coefficients controlling the degree of symmetry breaking
are assumed to be either explicit or to have an origin in spontaneous symmetry breaking.
The second version focuses on the weak-field regime where the spacetime metric can be expanded around a Minkowski metric as
\beq
g_\mn = \et_\mn + h_\mn, 
\label{metricexp}
\eeq
and uses a quadratic Lagrange density as the starting point.
In this case the coefficients are assumed to take their vacuum values and are coupled to the metric fluctuations $h_\mn$
in a way that is consistent with the spontaneous breaking of Lorentz symmetry.
Both of these approaches overlap in the linearized limit and we shall adopt the latter since our focus is on weak-field effects.

To start, 
note that the linearized field equations of General Relativity can be derived from a quadratic (in $h_\mn$) Lagrange density.
This takes the form
\beq
\cL_{GR} = - \frac {1}{4\ka} h^\mn G_\mn + \frac 12 h_\mn (T_M)^\mn,
\label{gr}
\eeq
where $G_\mn$ is the linearized Einstein tensor, 
we have included a conventional coupling to the matter stress-energy tensor $(T_M)^\mn$, 
and $\ka=8\pi G_N$.

In the context of quadratic actions and linearized field equations, 
the most general Lorentz and CPT-breaking action consistent with gauge symmetry 
(linearized diffeomorphism symmetry) is known \cite{bk06, bkx15, km16}.
The terms in this expression are organized by the mass dimension of the operator involving $h_\mn$ its derivatives,
with the Lagrange density for GR having a mass dimension of $4$ in natural units (or equivalently a length dimension of $4$).  
Coupled to these operators are the coefficients for Lorentz violation, 
which are labeled by the appropriate mass dimension of the operator.
The mass dimension $4$ term represents the lowest order, or minimal, 
Lorentz-breaking term that can be written in this series and it takes the form
\beq
\cL^{(4)}= \frac {1}{4\ka} \sb^{\mu\ka} h^{\nu\la} \cG_{\mu\nu\ka\la}, 
\label{L4}
\eeq
where $\cG_{\mu\nu\ka\la}$ is the double dual of the linearized Riemann curvature tensor
and the $9$ {\it a priori} independent coefficients are contained in the symmetric traceless
$\sb_\mn$.
This term has been extensively studied and independent measurements now exist from a variety of tests, 
both terrestrial and space-based \cite{tables}. 
The best solar-system limits on the dimensionless $\sb_\mn$ coefficients are at the level of $10^{-8} - 10^{-11}$ 
from lunar laser ranging \cite{bourgoin1617}, 
while constraints inferred from distant cosmic rays reach $10^{-13} - 10^{-14}$ on these coefficients \cite{kt15}.
Note that while the $\sb_\mn$ coefficients affect the propagation of gravitational waves through the dispersion relation, 
the resulting constraints from the observation of gravitational wave events \cite{gw} yield poor sensitivity compared
to those obtained by other tests \cite{km16}.
This result is in contrast to the higher mass dimension coefficients in the SME expansion.

While the minimal SME in the gravity sector has been explored, 
higher mass dimension terms in the Lagrangian have only begun to be explored.
The next two terms in the nonminimal SME expansion in the gravity sector can be written in terms of a covariant action \cite{bkx15}, 
or a quadratic effective action \cite{km16}.
For the latter form, 
the mass dimension $5$ operator term appearing in the SME expansion can be written as
\beq
\cL^{(5)}= -\frac {1}{16\ka} h_\mn (q^{(5)})^{\mu\rh\al\nu\be\si\ga} \prt_\be R_{\rh\al\si\ga}, 
\label{L5}
\eeq
where the coefficients are $(q^{(5)})^{\mu\rh\al\nu\be\si\ga}$ and have dimensions of inverse mass or length, 
and $R_{\rh\al\si\ga}$ is the linearized Riemann curvature tensor.
There is complete antisymmetry in the first $3$ indices and Riemann symmetry in the last four indices.
Using Young tableaux it can be established that there are $60$ independent coefficients.
The Lagrangian \rf{L5} breaks CPT symmetry for gravity, 
which is defined operationally as resulting from the operator $\prt_\be R_{\rh\al\si\ga}$
having an odd number of spacetime indices \cite{k04}.

It is important to note that the terms present in \rf{L5} are interpreted perturbatively, 
as small corrections to the dynamics of $h_\mn$ from GR.
This means that we do not consider modes in the associated dispersion relation
of higher than the second power in momentum, 
which essentially means we are avoiding Ostrogradski instabilities \cite{ost, km09}.
Note also that the condition on the partial derivatives of the coefficients for Lorentz violation, e.g., 
$\prt_\la \sb_\mn=0$, 
is assumed to hold throughout the analysis in this work in a suitable Cartesian coordinate system.
This point is discussed in more detail elsewhere \cite{k04,bk06}.

Beyond mass dimension $5$ are the coefficients for mass dimension $6$ operators.
These produce effects in short-range gravity tests which offer some of the best sensitivity \cite{SMEsr}, 
as well producing effects in gravitational wave propagation \cite{km16}.
In this work our focus is on the mass dimension $4$ and $5$ coefficients
and we leave it as an open question to determine the additional post-Newtonian effects of 
the mass dimension $6$ coefficients beyond the Newtonian limit.

Various models that exist in the literature can be directly matched to the Lagrange densities above.
This includes vector field models with a potential term driving spontaneous breaking of Lorentz and diffeomorphism symmetry
\cite{ks89}.
In particular,
some vector models considered in the literature include additional kinetic terms beyond the Maxwell one \cite{models}.
With certain constraints on these models they match the form of \rf{L4} once the dynamics
of the vector field have been imposed on the effective action \cite{ms09}.
Furthermore, 
models of spontaneous Lorentz-symmetry breaking with anti-symmetric and symmetric tensors
also can match the form of \rf{L4} \cite{models2}.
In fact, 
this is a general feature of the SME, 
whereby specific models can be matched to specific SME coefficients and
the existing limits can then be used to constrain them \cite{ms10}.
As further examples, 
matches to the SME exist with noncommutative geometry and quantum gravity \cite{match, km09}.

In the gravity sector, 
the field equations in the linearized limit, 
stemming from the combined Lagrange densities \rf{gr}, 
\rf{L4}, 
and \rf{L5} can be derived by varying with respect to the metric fluctuations $h_\mn$.
The result is
\beq
G^\mn = \ka (T_M)^\mn + \sb_{\ka\la} \cG^{\mu\ka\nu\la} 
- \frac 14 q^{\rh\al(\mu\nu)\be\si\ga} \prt_\be R_{\rh\al\si\ga},
\label{FE}
\eeq
which can then be used to solve for the post-Newtonian metric.
Note that we have abbreviated the dimensional superscript $(q^{5}) \rightarrow q$
to simplify expressions.

The field equations \rf{FE} satisfy the conservation laws associated with the linearized diffeomorphism symmetry
present in the Lagrange densities \rf{gr}, \rf{L4}, and \rf{L5}.
This implies the vanishing divergence of the Lorentz and CPT breaking terms on the right-hand side, 
which can be checked directly.
This is consistent with the linearized Bianchi identities and the conservation of the matter stress-energy tensor $T_M^\mn$.
Since the conservation laws hold, 
the origin of the coefficients is then compatible with the case of 
spontaneous breaking of Lorentz and diffeomorphism symmetry \cite{spont}.
In particular, 
the imposition of linearized diffeomorphism symmetry limits the possible forms of the Lagrange densities above.
For example, 
consider the construction of a mass dimension $4$ term of the form \rf{L4} with the coefficients $\tb_{\mu\nu\ka\la}$, 
having the symmetries of the Weyl tensor.
This fails to produce a nonvanishing term that is not a total derivative, 
when the constraint of linearized diffeomorphism symmetry is imposed, 
as explained in more detail in Refs.\  \cite{bk06, km16, bonder}.
Nonetheless, 
a term of this type may have consequences in cosmological scenarios \cite{bonder17}, 
which we do not explore in this work.

\section{Post-Newtonian expansion}
\label{pnexpansion}

We adopt standard assumptions for weak-field slow motion gravity to calculate the relevant post-Newtonian metric.
The perfect fluid stress-energy tensor is assumed for matter, 
and the Newtonian potential $U$ dominates as usual in this approximation method.
The field equations \rf{FE} are solved by decomposition into the space and time components 
and using successive corrections in powers of the small velocity $v$, 
which is assumed much less than unity.
These standard methods have been applied to the SME and the details are explained elsewhere \cite{bk06, b16}.
Note also that some of the terminology used here, 
like viewing velocity and acceleration as spatial ``vectors", 
is only valid in the post-Newtonian weak-field limit up to a certain ``order" in the expansion parameter $v$, 
and care is required in calculating observables and connecting them to real measurements.

In the results we record here we shall retain the post-Newtonian metric corrections up to order $v^3$ or 
post-Newtonian order $3$ ($PNO(3)$) to certain components of the metric, 
and up to $PNO(2)$ in other components.
We also include results for the CPT-even mass dimension $4$ coefficients $\sb_\mn$ for comparison 
to the CPT-odd mass dimension $5$ coefficients $q$.
Our coordinate choice is consistent with the harmonic gauge to the necessary post-Newtonian order.
To derive the metric, 
we make use of the following ``superpotentials" \cite{tegp, pw14}:
\bea
\ch &=& -G_N \int d^3r^\prime \rh^\prime |\vec r - \vec r^\prime|,
\nn\\
\ch^j &=&  G_N \int d^3r^\prime \rh^\prime v^{\prime j} |\vec r - \vec r^\prime|,
\label{super}
\eea 
and we solve for the metric to leading order in the coefficients.
In a space and time decomposition, 
the components of the metric are given by
\bea
g_{00} &=& -1+ 2U(1+3\sb_{00})+\sb_{jk} U^{jk} + 4 \sb_{0j} V^j + {\hat Q}^j \ch^j,
\nn\\
g_{0j} &=& \sb_{0k} (U^{jk} + \de^{jk} U) + \frac 12 {\hat Q}^j \ch+...,
\nn\\
g_{jk} &=& \de_{jk} [1+ (2-\sb_{00})U +\sb_{lm}U^{lm}] - \sb_{lj}U^{kl}- \sb_{lk}U^{jl} 
\nn\\
&&+2\sb_{00}U^{jk}+ {\hat Q}^{jk}\ch, 
\label{metric}
\eea
where $U^{jk}=\prt_{jk}\ch +\de_{jk} U$ and $V^j=(1/2){\bf \nabla}^2 \ch^j$.
The results are compactly displayed in terms of the derivative operators ${\hat Q}^j$ and ${\hat Q}^{jk}$.
These are given in terms of the underlying coefficients $q$ by
\bea
{\hat Q}^j &=& [q^{0jk0l0m}+q^{n0knljm} +q^{njknl0m}] \prt_{klm},
\nn\\
{\hat Q}^{jk} &=& [q^{0l(jk)m0n}+q^{pl(jk)mpn} +\de^{jk} q^{0pl0mpn} ] \prt_{lmn}.\nn\\
\label{hats}
\eea
The ellipses in $g_{0j}$ stand for $PNO(3)$ terms that are omitted for space since they are 
not needed for the analysis in this work.

The partially symmetrized combinations of coefficients in \rf{hats} occur frequently in what follows so we 
define effective coefficients combinations $K_{jklm}$ and ${\tilde K}_{jklmn}$ as follows:
\bea
K_{jklm} &=& -\frac 16 ( q_{0jk0l0m} + q_{n0knljm} + q_{njknl0m} 
\nn\\
&&
\pt{ -\frac 16  }
+ {\rm perms} ),
\nn\\
{\tilde K}_{jklmn} &=& \frac 16 ( q_{0l(jk)m0n}+q_{pl(jk)mpn} +\de_{jk} q_{0pl0mpn} 
\nn\\
&&
\pt{ -\frac 16  }
+ {\rm perms} ),
\label{Keff}
\eea
where perms indicates all symmetric permutations in the last three indices $klm$
and $lmn$, respectively, 
and we have lowered the indices with the Minkowski metric $\et_\mn$.
Certain properties of these coefficient combinations also hold
which are useful to simplify calculations.
For example, 
for any spatial vectors ${\vec a}$, $\vec b$, and ${\vec c}$ 
the following identities hold:
\bea
K_{jklm} a^j a^k a^l a^m &=& 0, 
\nn\\
K_{jklm} a^j a^k b^l b^m &=& -K_{jklm} b^j b^k a^l a^m,
\nn\\
K_{jklm} b^j a^k b^l b^m &=& -\frac 13 K_{jklm} a^j b^k b^l b^m.
\label{Kidents}
\eea

While the post-Newtonian metric contains all nine coefficients in $\sb_\mn$, 
only a subset of the $60$ {\it a priori} independent coefficients $q^{\rh\al\mu\nu\be\si\ga}$ appear. 
This implies that via post-Newtonian tests, 
not all of the mass dimension $5$ coefficients can be probed.  
Similar results hold for Lorentz-violating effects on gravitational wave propagation, 
where a subset of $16$ of these coefficients appear at leading order \cite{km16}.
The coefficients appearing in \rf{Keff} are combinations of the space and time decomposed 
irreducible pieces of the $q^{\rh\al\mu\nu\be\si\ga}$ coefficients. 
The combinations $\K5$ include the $15$ dimensional piece $q_{0jk0l0m}$, 
the $10$ dimensional piece $q_{0jklmnp}$, 
and the $8$ dimensional piece $q_{jklmn0p}$.
However, 
due to the symmetry properties of $\K5$ and \rf{Kidents}, 
there are only $15$ independent combinations of these irreducible pieces appearing.
Furthermore, 
only a subset of those will actually appear for a given experiment or observational analysis.
Similar considerations hold for the ${\tilde K}_{jklmn}$ combinations.

The equations of motion for self-gravitating bodies can be derived
from the metric components \rf{metric} and the standard fluid
equations contained in the conservation law $D_\mu (T_M)^\mn=0$.
This method uses a perfect fluid model for matter, 
as done previously for the SME in Refs.\ \cite{bk06} and \cite{b16}.
While we do not discuss it here, 
these methods can be generalized to the case of Lorentz violation in the matter sector
including gravitational effects \cite{kt09, kt11}.

We restrict attention to the case of two pointlike bodies with masses $m_a$ and $m_b$ 
and positions $\vec r_a$ and $\vec r_b$ in an asymptotically inertial coordinate system in which the 
coefficients are assumed constant \cite{bk06}.
The relative position between the two bodies is $\vec r = \vec r_a - \vec r_b$
and ${\hat n}=\vec r / r$ is a unit vector pointing in this direction,
while the relative velocity is $\vec v =  \vec v_a - \vec v_b$.
By suitably integrating the fluid equations over body $a$, 
the acceleration of body $a$ due to body $b$ is found to be
\bea
\fr {d^2 r_a^j}{dt^2} &=& -\fr {G_N m_b}{r^2} 
[ (1+\frac 32 \sb_{00} ) n^j - \sb_{jk} n^k  + \frac 32 \sb_{kl} n^k n^l n^j ]
\nn\\
&&
+\fr {2 G_N m_b}{r^2} (\sb_{0j}v^k n^k - \sb_{0k}v^k n^j)
\nn\\
&&
+\fr {G_N m_b}{r^2} \sb_{0k} v_b^l (2\de^{j(k} n^{l)} -3 \de^{kl} n^j -3 n^j n^k n^l)
\nn\\
&&
+\fr {G_N m_b v^k}{r^3} 
\big( 
15 n^l n^m n^n n_{[ j } K_{k] lmn} 
\nn\\
&&
+  9 n^l n^m K_{[jk] lm}-9 n_{[j} K_{k] ll m} n^m -3 K_{[jk]ll} 
\big)+...
\nn\\
\label{accela}
\eea
The first term proportional to $n^j$ is the Newtonian acceleration, 
followed by the acceleration modifications from the $\sb$ coefficients.
The final terms controlled by the mass dimension $5$ coefficient combinations $\K5$
can be viewed, 
in the context of the post-Newtonian expansion, 
as the result of a nonstatic (velocity dependent) inverse cubic force between the masses $a$ and $b$, 
which is strikingly different from what occurs in GR and other Lorentz-breaking terms \cite{bkx15, km17}.

Also in Eq.\ \rf{accela} the ellipses stand for corrections from GR and higher terms 
in a post-Newtonian series.
Note that there are no self-acceleration terms present, 
which is consistent with the fact that the SME is based on an action principle
with energy and momentum conservation laws. %need to check momentum of two-body system!
In fact, 
the result \rf{accela} can be derived from a post-Newtonian series 
of the standard geodesic equation with the metric \rf{metric}.

The equations for the relative acceleration of two bodies, 
which is more closely related to what is actually observable, 
are straightforward to compute from \rf{accela}.
In fact the expression for the relative acceleration of bodies $a$ and $b$, 
$a^j=d^2r^j/dt^2$, can be obtained from the right-hand side of \rf{accela} 
with the replacement of $m_b \rightarrow M=m_a+m_b$, 
with the exception of the $\sb_{0j}$ terms.
These latter terms contain a dependence on the Newtonian
center of mass velocity via $m_a \vec v_a + m_b \vec v_b$ \cite{bk06}.

It is also useful to write down the effective two-body classical Lagrangian from which the
equations of motion can be derived.
Specifically, for two bodies, 
$a$ and $b$, 
we have 
\bea
L &=& \frac 12 (m_a v_a^2 + m_b v_b^2) 
\nn\\
&&
+ \fr {G_N m_a m_b}{r} \left( 1+ \frac 32 \sb_{00}+\frac 12 \sb_{jk} n^j n^k \right)
\nn\\
&&
+\fr {G_N m_a m_b}{2 r} 
\left( 3 \sb_{0j} (v_a^j + v_b^j) +\sb_{0j} n^j (v_a^k+v_b^k) n^k \right)
\nn\\
&&
-\fr {3 G_N m_a m_b}{2 r^2} v_{ab}^j (K_{jklm} n^k n^l n^m - K_{jkkl}n^l). 
\label{efflag}
\eea
One can see from this Lagrangian (or from the acceleration equations) 
a distinction between some of the $PNO(3)$ terms proportional to the velocities of the bodies. 
In the second line of the Lagrangian, 
terms depending on the velocity of the bodies relative to the background $\sb_{0j}$
are present.
In contrast, 
for the $\K5$ terms on the last line, 
the velocity dependence is only on the relative velocity of the two bodies 
$\vec v =  \vec v_a - \vec v_b$.
This has implications for the subtle observability of these coefficients, 
as is revealed by looking at specific tests.

\section{Tests}
\label{tests}

For the specific tests discussed below, 
we adopt, 
where possible, 
the standard Sun-centered Celestial Equatorial Frame coordinates (SCF) 
in which coefficient measurements are reported \cite{sunframe, tables}.
These coordinates are denoted with capital letters as $\{T,X,Y,Z \}$.
For convenience, 
some results are projected along various unit vectors associated with the specific test.
These unit vectors can then be expressed in terms of the SCF coordinates for analysis.
Also, 
the reader is cautioned about the notational changes in the following discussions where
the same symbols may be used to represent different quantities in different subsections.

We remark in passing that the tests discussed below are also of interest for the matter-gravity couplings
$\ab_\mu$ and $\cb_\mn$, 
and the associated phenomenology is published elsewhere \cite{kt09, kt11, mgcoup}.
However, 
the $\cb_\mn$ coefficients for the electron, proton, and neutron, 
while in principle measurable in gravity tests,
are now better constrained from laboratory and astrophysical tests \cite{tables, Cesium}.

It should also be emphasized that for the orbital tests discussed below, 
we use the point-mass approximation in Eq.\ \rf{accela} and \rf{efflag}.
While this suffices for bodies that are sufficiently separated, 
for cases such as near Earth satellites, 
it may be necessary to include effects from the bodies spherical inertia that
arise in the potentials \cite{bk06}.

\subsection{Secular changes}
\label{secular changes}

Using the method of oscillating orbital elements and the acceleration \rf{accela}, 
one can calculate the time derivatives of six Keplerian orbital elements for a generic binary orbit \cite{osc}.
Of primary interest for analysis are five of these elements: 
the semi-major axis $a$, 
the eccentricity $e$, 
the periastron $\om$, 
the inclination with respect to the chosen reference plane $i$, 
and the angle of the ascending node $\Om$.

After calculating the changes in the elements using the modified acceleration we then time average
the results over one orbital period, 
yielding the secular changes.
The formulas for the secular changes in the orbital elements due to the $\sb_\mn$ coefficients can 
be found in equations (168)-(171) in Ref.\ \cite{bk06} and elsewhere \cite{shao14, hees15}.
We find that the secular change in the semimajor axis and eccentricity both vanish for the $\K5$ coefficients.
This is in contrast to the coefficients $\sb_\mn$, 
where a contribution to the change in the shape of the orbit $e$ persists after averaging.
However, 
for the $\K5$ coefficients,
the orientation of the orbit changes via the periastron, 
inclination, 
and ascending node angle as follows:
\bea
\langle \fr{d \om}{dt} \rangle &=& -\frac {n^2}{4(1-e^2)^{3/2}}  
\{2 K_1 + \cot i [ \cos \om K_2 
\nn\\ &&\pt{more and more space} +\sin \om K_3] \}, \\
\langle \fr{di}{dt} \rangle &=& \frac {n^2}{4(1-e^2)^{3/2}}
[ \cos \om K_3 - \sin \om K_2 ], \\ 
\langle \fr{d\Om}{dt} \rangle &=& \frac {n^2}{4(1-e^2)^{3/2}}  \csc i
[ \cos \om K_2 +\sin \om K_3],
\label{secular}
\eea
where the combinations $K_1$, $K_2$, and $K_3$ are given by
\bea
K_1 &=& 3 K_{PPPQ}+ K_{PQQQ} + 6 K_{[PQ]kk},
\nn\\
K_2 &=& 3 K_{PQQk} - 3 K_{PPPk} - 4 K_{Pkkk} - 6 K_{QPQk},
\nn\\ 
K_3 &=& 6 K_{PPQk} + 4 K_{Qkkk}-3 K_{QPPk}+3 K_{QQQk}, \nn\\
\label{Kcombs}
\eea
and $n$ is the frequency of the orbit.
The subscripts on the coefficients stand for projections along the three unit vectors defining 
the orientation of each orbit: $\vec P$, $\vec Q$, and $\vec k$ (see Eq.\ (166)
in Ref.\ \cite{bk06}).

The results above indicate that three combinations of the coefficients $\K5$ 
appear in the secular changes of the $3$ orbital elements above.
The use of multiple orbits, 
each with differing orientation, 
can disentangle the coefficients.
Analyses similar to the study of solar-system ephemeris in Refs.\ \cite{hees15,iorio12} 
and binary pulsar observations in Ref.\ \cite{shao14} would be of interest.
In particular it should be noted that while the $\sb_\mn$ coefficients also yield 
changes in the elements $i$, 
$\om$, 
and $\Om$,
the angular dependence of the coefficient projections for the mass dimension $5$ coefficients, 
such as $K_{PQQk}=K_{jklm} P^j Q^k Q^l k^m$, 
differs.
This may be used to disentangle the $\K5$ coefficients from the $\sb_\mn$ coefficients.

The possibility also exists to study more extreme orbits.
For example, 
orbital speeds on the order of $10^{-3}c$ have been observed
in the stars near the center of our galaxy where evidence points to the existence
of a supermassive black hole, 
and this could make an interesting testing ground for CPT and Lorentz symmetry \cite{SMBH}.
This is particularly noteworthy since the modifications to the force between two point masses varies as the inverse cube of 
the distance for the $\K5$ coefficients.

\subsection{Laser ranging}
\label{laser ranging}

To obtain observable range oscillations for the lunar case and also Earth satellite case,
we expand perturbatively around a circular orbit using standard methods described in the literature \cite{tegp, lunarosc}.
Implicitly then our approximate results will be valid for nearly circular elliptical orbits.
The basic circular orbit frequency is $\om$ while the so-called ``anomalistic frequency" is $\om_0$.
The latter represents the frequency of the natural eccentric oscillations around the circular orbit and
the difference in the two frequencies reflects a perigee precession arising from large Newtonian perturbations.
For the present application, 
the equation of motion for the relative position of the satellite and source body is best expressed in the form
\beq
a^j = -\fr {GM r^j}{r^3} + \vec \nabla \de V (r) + \de a^j +...,
\label{accrange}
\eeq 
where $\de V (r)$ represent the ``central" portion of the perturbative potential (both Newtonian, Lorentz-violating, and otherwise) 
which is used in the definition of the circular orbit frequency $\om$.
Here $\de a^j$ is the Lorentz-violating acceleration, 
obtainable from \rf{accela}.

We then expand around a circular orbit radius $r_0$ as $r=r_0+\de r$ and similarly we expand the angular momentum per unit mass
$h=|\vec r \times \vec v |$ as $h=h_0 + \de h$ and truncate the result order by order assuming the perturbations are small.
The mean speed of the orbit is $v_0=\om r_0$.
The basic equation for the oscillations in the range $\de r$ between two bodies is given approximately, 
to second order in perturbations, 
by
\beq
\de \ddot{r} + \om_0^2 \de r = \fr {h_0 2 \de h}{r_0^3} + \fr {(\de h)^2}{r_0^3}+ \fr {3 \om^2 (\de r)^2}{r_0} 
- \fr {6 h_0 \de h \de r}{r_0^4} + \de a_r,
\label{deltarODE}
\eeq
where $\de a_r=\hat n \cdot \de \vec a$.
The second order terms are kept to include some near-resonant terms that arise from coupling Lorentz-violating oscillations 
with the basic eccentric oscillation
\beq
\de r_e = r_0 e \cos (\om_0 t+ \ph).
\label{ecc_osc}
\eeq

The equation \rf{deltarODE} is a driven harmonic system and we can catalog the dominant oscillations controlled by the 
coefficients $\K5$.  
To contrast the results with the minimal SME we also include the results for the $\sb_\mn$ coefficients.
Table \ref{rangeosc} lists the amplitudes and frequencies for the oscillation signal, 
which is described by the general form
\beq
\de r = \sum_n [A_n \cos (\om_n T + \phi_n) + B_n \sin (\om_n T + \phi_n)],
\label{delr}
\eeq
where $\om_n$ and $\phi_n$ are the frequencies and associated phases.
The details on the phases of the oscillations can be found in Ref.\ \cite{bk06}.
Also, 
the coefficients appearing in the amplitudes are displayed compactly in terms of projections onto an orbital plane basis
$\{ {\bf e}_1, {\bf e}_2, {\bf e}_3 \}$, 
for example, 
$K_{1122}=K_{JKLM} e_1^J e_1^K e_2^L e_2^M$.
This basis can be expressed in terms of the SCF coordinates using the results in Ref.\  \cite{bk06}, 
in particular the satellite orbit figure $4$ of that reference.

\begin{table}
\caption{\label{rangeosc}
Dominant range oscillation frequencies and amplitudes for the lunar and satellite laser ranging scenario.
We include mass dimension $4$ coefficients $\sb_\mn$ and the CPT-violating mass dimension $5$ coefficients $\K5$.
The coefficients are projected onto an orbital plane basis.}
\begin{ruledtabular}
\begin{tabular}{lcc}
Frequency & $\sb_\mn$ Amplitude & $ \K5 $ Amplitude \\ 
\hline
$A_{2\om}$ & $-\frac {1}{12}r_0(\sb_{11}-\sb_{22})$ & $\frac 14 v_0 (6 K_{(12)33}-3 K_{1112}+K_{1222})$ \\
$B_{2\om}$ & $-\frac 16 r_0 \sb_{12} $ &  $-\frac 34 v_0 (K_{1122}+K_{1133}-K_{2233})$ \\
$A_{2\om-\om_0}$ & $ -\frac {\om e r_0 (\sb_{11} -\sb_{22})}{16(\om-\om_0)}  $ & 0 \\
$B_{2\om-\om_0}$ & $- \frac{\om e r_0 \sb_{12}}{8(\om-\om_0)} $  & 0 \\
$A_{\om}$ & $\frac {\om v_0 r_0 \sb_{02}}{(\om-\om_0)}$ & 0\\
$B_{\om}$ & $-\frac{\om v_0 r_0 \sb_{01}}{(\om-\om_0)}$ & 0 \\
\end{tabular}
\end{ruledtabular}
\end{table}

The oscillations for the near resonance frequencies $\om$ and $2\om-\om_0$ are absent for the $\K5$ coefficients.
Furthermore, 
as displayed in Ref.\ \cite{bk06}, 
the $\sb_{0j}$ coefficients lead to oscillation also at the Earth's orbital frequency $\Om$, 
while no such terms arise for the $\K5$ coefficients.
The latter result can be traced to the exclusive relative velocity dependence in \rf{accela}.
Higher harmonics in the frequencies $\om$ and $\Om$ also exist but are typically suppressed 
by $e$ or powers of the factor $\om/\Om$ relative to the ones appearing in Table \ref{rangeosc}.

While post-fit analysis using Eq.\ \rf{delr} can be performed \cite{battat07}, 
a more rigorous analysis includes the equations of motion \rf{accela} directly into the ephemeris code for laser ranging observations.  
Such an analysis has been performed recently for the $\sb_\mn$ coefficients and has placed the most stringent solar-system 
limits on these coefficients using decades of data from lunar laser ranging \cite{bourgoin1617}.
It would be of definite interest to try to add the modifications for the $\K5$ combinations of the $q$ coefficients into this code
and perform a combined fit with the $\sb_\mn$ coefficients.
As the table indicates, 
there is likely to be significant correlation of the $\K5$ coefficients with the $\sb_\mn$ coefficients.
Though it appears to represent a challenge to disentangle them,
a full analysis may reveal signals from the $\K5$ coefficients that are distinct from the $\sb_\mn$ coefficients.
Also, 
since orbits of differing orientation can help disentangle the coefficients, 
it would also be of interest to perform analysis with satellite orbits \cite{sat}.

\subsection{Light propagation}
\label{light}

Among other precision tests in the solar system is the observation of the deflection of light around massive bodies. 
This can be used to test spacetime symmetry via the effects of the metric in \rf{metric}.
One particularly useful relativistic effect is the time delay of light \cite{shapiro64}.
We present here the one way coordinate time difference between event $e$ (emission) and event $p$ (reception).
The basic setup for this problem in any weak-field metric based theory of gravity is discussed in generality in Ref.\ \cite{b09}.
We employ the quantities defined in table \ref{tdquant} for a straight line trajectory between the emission and reception events:

\begin{table}
\caption{\label{tdquant}
Quantities used for the time delay formula \rf{td2}.  The reader is referred to figure 1 in Ref. \cite{b09}.}
\begin{ruledtabular}
\begin{tabular}{cc}
Quantity & Definition  \\ 
\hline
$\vec r_p$ & Spatial coordinate of event $p$ \\
$\vec r_e$ & Spatial coordinate of event $e$ \\
$\vec R$ & $\vec r_p - \vec r_e$ (unperturbed displacement) \\
$\hr$ & Unit vector in the direction of $\vec R$ \\
$l_e$ & $-\vec r_e \cdot \hr$ \\
$l_p$ & $\vec r_p \cdot \hr $ \\
$\vec b$ & $\vec r_p - l_p \hr$ (impact parameter vector) \\
\end{tabular}
\end{ruledtabular}
\end{table}

The coordinate time difference is obtained by integration of the metric fluctuations 
projected and evaluated along the unperturbed straight path, 
as presented in equation (12) of Ref.\ \cite{b09}:
\beq
t_p - t_e = R + \fr 12 \int_{-l_e}^{l_p} h_\mn \pb^\mu \pb^\nu d\la.
\label{td1}
\eeq
In this expression, 
$\la$ is a parameter in the unperturbed trajectory $x_0^j = \hr^j \la + b^j$ which is inserted into the metric
while $\pb^0 =1$ and $\pb^j=\hr^j$.

After integration with the metric \rf{metric},
we obtain
\begin{widetext}
\bea
t_p-t_e &=& R +
GM \bigg\{ 2 (1+\sb_{00}+\sb_{0j} \hat R^j ) \ln \left[\fr {r_e + r_p + R}{r_e +r_p -R}\right] 
-[\sb_{00}+ \sb_{0j} \hat R^j-\sb_{jk} \hb^j \hb^k] \left(\fr {l_e}{r_e} +\fr {l_p}{r_p}\right) 
\nn\\
&&
- [\sb_{0j} b^j + \sb_{jk} \hat R^j b^k] \fr {(r_e - r_p)}{r_e r_p} 
+T_{lmn} \hr^l \hr^m \hr^n b^2 \left( \fr {1}{r_e^3}-\fr {1}{r_p^3} \right) 
+T_{lmn} \hr^l \hr^m \hb^n \fr {3b}{2} \left( \fr {l_e}{r_e^3} + \fr {l_p}{r_p^3} \right) 
\nn\\
&&
+T_{lmn} \hr^l \hb^m \hb^n \fr 32 \left[ \fr {1}{r_e} - \fr {1}{r_p}+b^2 \left( \fr {1}{r_p^3} - \fr {1}{r_e^3} \right) \right]
+T_{lmn} \hr^l {\hat \ta}^m {\hat \ta}^n \fr 32 \left( \fr {1}{r_e} - \fr {1}{r_p} \right)
\nn\\
&&
+T_{lmn} \hb^l \hb^m \hb^n \fr {3}{2b} \left( \fr {l_e^3}{r_e^3} + \fr {l_p^3}{r_p^3} \right) 
+T_{lmn} \hb^l {\hat \ta}^m {\hat \ta}^n \fr {3}{2b} \left( \fr {l_e}{r_e} + \fr {l_p}{r_p} \right) 
\bigg\},
\label{td2}
\eea
\end{widetext}
where $T_{lmn}$ is given by
\beq
T_{lmn} = K_{jlmn}{\hat R}^j + {\tilde K}_{jklmn} \hr^j \hr^k.
\label{T}
\eeq
Here the unit vector ${\hat \ta}$ is defined by ${\hat \ta}=\hr \times \hb$.

The results for the $\sb_\mn$ coefficients were obtained previously and have been used 
to constrain the isotropic coefficient $\sb_{TT}$ at the level of $10^{-4}$ from 
Very Long Baseline Interferometry measurements \cite{VLBI}.
While these measurement are not as competitive with other tests for the $\sb_\mn$ coefficients, 
possible future analysis could now include the mass dimension $5$ coefficient combinations 
contained in \rf{T}.
Through the unit vectors $\hat R$ and $\hat b$, 
the time delay measured depends on the orientation of the receiver and the 
massive body $M$, 
which will typically change as a function of time throughout the observation period.
It would be of interest to perform an analysis to search for the CPT-violating coefficients
using time delay measurements.

Through the metric, 
light propagation can be affected in other ways.
The standard gravitational redshift, 
and the bending of light would be affected by the $q$ coefficients as well, 
which could be of interest for other tests \cite{tb11, gravexpt}.

\subsection{Earth laboratory tests}
\label{lab tests}

Among the more sensitive probes of gravity are laboratory tests on Earth involving
measurements of the free fall of masses near the surface, or gravimeter tests.
The instruments used include superconducting spheres suspended electromagnetically 
and measurements of the free fall acceleration via atom interferometry \cite{wg76,AI}.

In the Earth laboratory setting, 
the locally measured free fall acceleration of a test body
will be modified by the coefficients for Lorentz violation.
In the case of the mass dimension $4$ coefficients $\sb_\mn$, 
the dominant effects are oscillations in the free fall acceleration at different harmonics
of the Earth's sidereal frequency $\om$ and the Earth's orbital frequency $\Om$.
The amplitudes in terms of the SCF coefficients are tabulated in Ref.\ \cite{bk06}.
We record below the main results for the mass dimension $5$ coefficients $\K5$ which
affect the free fall motion of masses on the Earth's surface. 
Some the results for the $\sb_\mn$ coefficients are retained for comparison.
Note that for simplicity we do not include here the effects of the Earth's finite size which is known 
to produce extra significant acceleration terms involving the Earth's spherical inertia \cite{bk06}.
These effects can readily be incorporated by using the super potentials \rf{super} 
in the metric \rf{metric}.

To obtain the local modified acceleration for the Earth laboratory setting one can proceed
from the effective Lagrangian in \rf{efflag}, 
treating body $a$ as the test mass and body $b$ as the source body (Earth).
One then expands around a point on the Earth's surface (assuming the axes
are oriented with $\hat z$ being the local vertical, 
$\hat y$ points east, 
and $\hat x$ points south).
Alternatively, 
one can use the local metric of an accelerated and rotating observer in a generic space-time
and proceed by calculating the local acceleration using a covariant expression, 
as done in Ref.\ \cite{bk06}.
The basic signal one obtains takes the form
\beq
\fr {\de a^{\hat z}}{a^{\hat z}} = \sum_n [C_n \cos (\om_n T + \phi_n) + S_n \sin (\om_n T + \phi_n)],
\label{dela}
\eeq
where the frequencies are labelled by $n$, 
$C_n$ and $S_n$ are the coefficient-dependent amplitudes and
$\ph_n$ are the associated phases. 
Note that the time $T$ is the SCF time and 
the phase is $\ph=\om (T_{\oplus} - T)$, %continuity of notation
where $T_{\oplus}$ is defined relative to the crossing of the local ${\hat y}$ axis 
with the Sun-centered frame $Y$ axis \cite{sunframe}.
The amplitudes for the frequencies $\om$, 
$2\om$, 
and $3\om$ are given by
\begin{widetext}
\bea
C_\om &=& -\frac 12 \sin 2\ch \sb_{XZ} -2 \om R_{\oplus} \sin \ch \sb_{TY}  
+ \frac { \om }{32} [ 
K_{XXYZ} (30 +18 \cos 2\ch ) 
+K_{YXXZ} (3 - 27 \cos 2\ch ) 
\nn\\
&&
\pt{-\frac 12 \sin 2\ch \sb_{XZ} -2 V_L \sb_{TY} +}
+K_{YYYZ} (33 - 9 \cos 2\ch ) 
+K_{YZZZ} (-28 + 12 \cos 2\ch ) 
] \sin 2 \ch,
\nn\\
S_\om &=& -\frac 12 \sin 2\ch \sb_{YZ} + 2 \om R_{\oplus} \sin \ch \sb_{TX} 
+ \frac { \om }{32} [ 
K_{XXXZ} (-33 +9 \cos 2\ch ) 
+K_{XYYZ} (-3 +27 \cos 2\ch ) 
\nn\\
&&
\pt{\frac 12 \sin 2\ch \sb_{YZ} + 2 V_L \sb_{TX} }
+K_{XZZZ} (28 - 12 \cos 2\ch ) 
-K_{YXYZ} (30 +18 \cos 2\ch ) 
] \sin 2 \ch,
\nn\\
C_{2\om} &=& - \frac 14 \sin^2 \ch (\sb_{XX}-\sb_{YY})
+ \frac { 3 \om }{8} [ 
(3 K_{XXXY}-K_{XYYY})(3+\cos 2\ch )
+3 ( K_{XYZZ}+K_{YXZZ} ) (\cos 2 \ch -1) 
] \sin^2 \ch,
\nn\\
S_{2\om} &=& -\frac 12 \sin^2 \ch \sb_{XY}
+ \frac {9 \om}{8} [
K_{XXYY} (3 + \cos 2 \ch )
+ (K_{XXZZ}-K_{YYZZ})(1 - \cos 2 \ch ) 
] \sin^2 \ch,
\nn\\
C_{3 \om} &=& \frac {9 \om}{8} (2 K_{XXYZ} + K_{YXXZ} - K_{YYYZ} ) \cos \ch \sin^3 \ch,
\nn\\
S_{3 \om} &=& -\frac {9 \om}{8} ( K_{XXXZ} - K_{XYYZ} - 2 K_{YXYZ} ) \cos \ch \sin^3 \ch,
\label{delaAMPS}
\eea
\end{widetext}
where $\ch$ is the experiment colatitude, 
$R_{\oplus}$ is the Earth's radius, 
and the phases for these frequencies are
$\ph$, 
$2 \ph$, 
and $3\ph$, 
respectively \cite{bk06}.

Note the appearance of the Earth sidereal rotational frequency $\om$ which provides
a dimensional quantity setting the scale for the sensitivity to the $\K5$ coefficients, 
which themselves have dimensions of time or length in natural units.
There is no leading order dependence of the signal on the Earth's orbital velocity
for the $\K5$ coefficients.
Therefore the harmonics that occur for the $\K5$ coefficients are simply multiples 
of the Earth sidereal rotational frequency, 
unlike the case of the $\sb_\mn$ coefficients, 
for which the subset $\sb_{0j}$ also appears at harmonics of Earth's orbital frequency $\Om$.
This can again be traced to the dependence on only relative velocity in the Lagrangian \rf{efflag}.
One distinction with the mass dimension $5$ coefficients is the appearance of the frequency $3\om$, 
which provides a way to disentangle these sets of coefficients.
Otherwise, 
a signal for CPT violation in gravimeter tests is entangled with the CPT-even effects of $\sb_\mn$.
While it appears to be challenging, 
it would be of definite interest to search for the $q^{\mu\rh\al\nu\be\si\ga}$ coefficients
in Earth-laboratory gravimeter tests of all types \cite{gravimeter, flowers16}.

Short-range gravity experiments, 
in which the force between two laboratory masses is carefully measured
as a function of separation, 
has been a useful probe of the SME gravity sector.
Results already limiting combinations of the mass dimension $6$ coefficients 
have been published \cite{SMEsr}.
In that case, 
the Newtonian potential is directly affected \cite{bkx15}.
However, 
for the coefficients $q^{\mu\rh\al\nu\be\si\ga}$, 
the contribution to the Newtonian potential vanishes \cite{km17}.
The leading order effects for these coefficients arise from terms
dependent on the velocity of the bodies as seen from the point-mass equations \rf{accela}.
For typical laboratory mass velocities, 
this introduces a suppression factor, 
making these tests less sensitive than others.
However, 
optimization may surmount this difficulty and it may be of interest also 
to investigate short-range tests.
One could proceed, 
in this case, 
by numerically integrating the point-mass acceleration formula
over the source and test bodies.
Note that a dependence on the motion of the source and test bodies is present
and this must be accounted for in the integration process.

\section{Discussion and Summary}
\label{discussion}

In this work, 
we investigated the gravity sector of the SME framework in the linearized gravity limit.
In particular, 
the post-Newtonian phenomenology of the mass dimension $5$ spacetime-symmetry breaking terms 
was studied.
The basic action is described in Section \ref{theory}, 
where the mass dimension $5$ term is given in equation \rf{L5} and the modified 
field equations are in equation \ref{FE}. 
The CPT-violating effects are controlled by the $60$ $q^{\mu\rh\al\nu\be\si\ga}$ coefficients
and the results in this work were contrasted with the mass dimension $4$ coefficients $\sb_\mn$, 
and other work on mass dimension $6$ coefficients and higher-order terms.

While previous studies of these coefficients focused on limits from recent gravitational wave events \cite{km16}, 
we focused in this work on the merits of weak-field slow motion gravity tests like those in the solar system 
or the Earth laboratory setting.
One key result in this work is the post-Newtonian metric, derived in Section \ref{pnexpansion} and
displayed in equation \rf{metric}.
This result shows the dominant modifications of GR controlled by the $q^{\mu\rh\al\nu\be\si\ga}$ coefficients.
Also in this section,  
the point-mass equations of motion for a binary system are derived.
The result for a body $a$ in the presence of body $b$ is 
given in Eq.\ \rf{accela} and displays a nonstatic inverse cubic distance behavior.

In Section \ref{tests}, 
we used the results of the previous sections to study particular solar system tests of gravity, 
discussing their merits as sensitive probes of spacetime symmetry violation in gravity.
The main results for orbital tests include formulas for the secular changes of orbital elements in Eq.\ \rf{secular}
and the dominant range oscillations for lunar and satellite laser ranging in equation \rf{delr} and Table \ref{rangeosc}.
The modified time delay formula is presented in equation \rf{td2}, 
and the oscillation amplitudes for the locally measured value of freefall acceleration are
contained in equations \rf{delaAMPS}.

To summarize the phenomenology,
we include here a table estimating the benchmark sensitivity 
for the different types of tests discussed in this work for measuring
combinations of the $q^{\mu\rh\al\nu\be\si\ga}$ coefficients, 
where the units of these coefficients are taken in kilometers.
Most of the tests discussed in this work depend on the $\K5$ combinations of the underlying
coefficients in the Lagrange density,
defined in \rf{Keff}.
The basic sensitivity levels can be 
established from the results derived in this paper, 
and those already known for the $\sb_\mn$ coefficients. 
For instance, 
examination of the equations of motion for two bodies in \rf{accela}, 
comparing the dimensionality of the $\sb_\mn$ and $\K5$ terms,
we can see that they are roughly related by a factor of velocity $v$ and
inverse distance $r$.
Since most tests involve a kind of cyclical motion, 
we arrive at the heuristic formula $K \sim \sb /\om$,
where $\om$ is a characteristic frequency of the system under study.
This works in most of the cases considered here,
as can be confirmed by examining the more experiment or test-specific results
in the paper (e.g., Table \ref{rangeosc}).
We can then extract the approximate sensitivity for the $\K5$ coefficients 
from the known limits on the $\sb_\mn$ coefficients.
For example, 
for solar system ephemeris, 
the limits on $\sb_{jk}$ coefficients are on the order of $10^{-10}$.
To find an estimate for what will be obtained for $\K5$ coefficients
we multiply $10^{-10}$ by $c / \om$, 
where $\om$ is the orbital frequency of the Earth and $c$ is the speed of light in SI units, 
thereby obtaining $K \sim 10^2 \, {\rm km}$.

The dependence of the sensitivity on frequency implies that satellite orbit analysis may 
result in increased sensitivity, 
as indicated in the Table.
For short-range gravity tests we can compare to the 
level at which the Newtonian gravitational force can be measured in a given experiment
and the frequency of motion of the masses in the experiment.
It is important to note that the sensitivity estimates provided 
do not address the issue of disentangling the $\K5$ coefficients from
the $\sb_\mn$ coefficients as discussed throughout
this work but merely provides a rough guide for analysis.
Also, 
while we do not investigate it here, 
the class of experiments involving gyroscopic precession, 
such as Gravity Probe B, 
could also be of potential interest for the $\K5$ coefficients \cite{gyro}.

\begin{table}
\caption{\label{estimates} Estimated sensitivity levels for different tests for the $\sb_\mn$ coefficients 
and the CPT-violating $q$ coefficients.  
The combinations
$\K5$ are defined in Eq.\ \rf{Keff}.  
Time delay tests are sensitive to both the $\K5$ coefficients and the ${\tilde K}_{jklmn}$ coefficients 
via the combination $T_{lmn}$ in Eq.\ \rf{T}.  
The values for $\sb_{JK}$ come from known limits on the spatial coefficients in the SCF \cite{tables}.
The range of values for the laser ranging case is due to the higher frequency of Earth satellite orbits.}
\begin{ruledtabular}
\begin{tabular}{ccc}
Test & $\sb_{JK}$ & $\K5$ (km) \\
\hline
solar system ephemeris & $10^{-10}$ & $10^2$ \\
binary pulsars & $10^{-10}$ & $10^{-1}$ \\
laser ranging & $10^{-11}$ & $10-10^{-1}$ \\
gravimeter & $10^{-9}$ & $5$ \\
short-range gravity & $10^4$ & $ 10^{4}$ \\
time delay & $10^{-4}$ & $10^{3}$ \\
\end{tabular}
\end{ruledtabular}
\end{table}

Finally we remark here about the possible sizes of the coefficients $q^{\mu\rh\al\nu\be\si\ga}$ discussed in this work.
A broad class of possible effects is described by the SME effective field theory framework.
As a test framework, 
there are no specific predictions concerning the sizes of the coefficients.
Nonetheless, 
since gravity is weak compared to other forces in nature, 
this leaves room for violations of spacetime symmetry that are large compared to those in other sectors, 
as evidenced by glancing at Table \ref{estimates} and comparing to the Planck length $10^{-35} \, {\rm m}$.
It is clear from this that symmetry breaking effects that are not ``Planck suppressed" could still have escaped detection.  
This effect is called ``countershading" and was first pointed out for matter-gravity couplings \cite{kt09} but persists also
in the higher mass dimension terms of the nonminimal SME \cite{b16}.
Although a more stringent limits exists on one combination of the mass dimension $5$ coefficients at the level of $10^{-14}$ m \cite{km16},
large, 
as yet unmeasured, 
CPT and Lorentz violation could still persist in nature.

\begin{acknowledgements}

We thank Adrien Bourgoin, Christine Guerlin, Aur\'elien Hees, V.A.\ Kosteleck\'y, 
Christophe Le Poncin-Lafitte, Jay Tasson, 
and an anonymous referee for valuable comments on the manuscript.  
This work was supported in part by the National Science Foundation under grant number PHY-1402890.

\end{acknowledgements}

\end{document}